\documentclass[%
 reprint,
 amsmath,amssymb,
 aps,
]{revtex4-2}

\usepackage{graphicx}% Include figure files
\usepackage{dcolumn}% Align table columns on decimal point
\usepackage{bm}% bold math
\usepackage{amsmath,amssymb}
\usepackage{physics}
\usepackage{xcolor}
\usepackage{soul}
\usepackage{layout}
\usepackage{hyperref}
\usepackage{cleveref}
\usepackage{float}
\usepackage{subcaption}

\begin{document}

\preprint{APS/123-QED}

%\title{Numerical test of the Generalized Langevin Equation with a potential of mean force and a linear memory kernel}
\title{How wrong is too wrong: A numerical study on the relevance of positional memory in the generalized Langevin equation}

\author{Abhir Mehrotra}
\author{Fabian Koch}
\author{Tanja Schilling}
 \email{Tanja.Schilling@physik.uni-freiburg.de}
\affiliation{%
 Institut f\"ur Physik, Albert-Ludwigs-Universit\"at Freiburg,\\ Hermann-Herder-Stra\ss e 3, 79104 Freiburg im Breisgau, Germany
}%

\date{\today}% It is always \today, today,

\begin{abstract}
 If a generalized Langevin equation contains a potential of mean force, it cannot at the same time contain a linear memory kernel and a fluctuating force that obeys a second fluctuation dissipation theorem in the sense of Kubo, and be exact. As modelers often prefer to use generalized Langevin equations that have the first three properties, one needs to ask how close the model dynamics is to the dynamics of the underlying microscopic system. To test this, we analyze a simple model system in which the potential of mean force can be well approximated by a polynomial of low order. The exact generalized Langevin equation of this model contains memory terms in addition to the linear one.
 We show that these additional terms, at least for the model system regarded in this article, are important for the dynamics and cannot be neglected if one intends to model core aspects of the underlying system correctly. 
\end{abstract}

\maketitle

%\tableofcontents
\section{Introduction}
In soft matter physics and biomolecular modeling, evolution equations of the Langevin-type are often used as effective coarse-grained models \cite{snook2006,Schilling2022}. 
Such models can be deduced from the underlying microscopic dynamics by means of projection-operator techniques \cite{grabert2006projection,zwanzig2001nonequilibrium}. 
In recent years there has been an extensive discussion on the question, which forms of generalized Langevin equations can in principle be derived via the projection-operator formalism \cite{carof2014coarse,vroylandt2022position,vroylandt2022derivation,glatzel2022interplay,koch2026snap,ayaz2022self}. 

From the modeling perspective there are two key features that are of particular interest in this context: First, to allow for a  straight-forward interpretation of the generalized Langevin equation (GLE) as a stochastic differential equation, there needs to be a simple fluctuation-dissipation relation between the memory kernel and the fluctuating force. Second, the GLE should contain a thermodynamic force that produces the correct equilibrium distribution of the physical observable. It has been shown that these two features can only be satisfied at the same time for very specific model systems such as the harmonic-oscillator heat bath \cite{zwanzig2001nonequilibrium}. In general, a potential of mean force is accompanied by a memory term which is non-linear in the observable and a fluctuation dissipation relation that is considerably more complicated than Kubo's second fluctuation dissipation theorem \cite{glatzel2022interplay,koch2026snap,vroylandt2022derivation,ayaz2022self}. Nevertheless, models which have both properties are still often employed, in particular when Grote-Hynes theory \cite{grote1980stable} is used to analyze chemical reaction or biomolecular folding problems (see e.g.~\cite{chaudhury2006approximate,satija2019generalized,singh2021generalized,roy2024memory} for concrete applications and \cite{schmidt2015simulation,drozdov2001improved,lavacchi2022non,zhou2025rapid} for general theoretical analyses.)

The projection-operator proposed by Mori \cite{Mori_Transport_1965}  yields a simple fluctuation-dissipation relation, the so-called second fluctuation-dissipation theorem (2FDT) \cite{kubo1966fluctuation}. If one prefers to have a thermodynamic force in the GLE, one uses a projector of the type introduced by Zwanzig \cite{zwanzig1961} or Chorin \cite{chorin2002optimal}, i.e.~a conditional expectation of an observable in a thermodynamic ensemble. We note that this type of projection can be interpreted as an infinite-rank Mori projection, i.e.~a Mori projection onto infinitely many variables which form a basis of the space of observables. 

However, if the thermodynamic force can be expressed in terms of only few of the infinitely many basis functions, one may obtain a generalized Langevin equation with only few additional additive memory terms. These additional memory terms are convolutions with different powers of the position and, hence, we refer to them as positional memory. (Note that this is different from the notion of position-dependent memory kernels in ref.~\cite{vroylandt2022position,ayaz2022self}.) In this article we analyze a model that is designed to fall exactly into this category. This allows us to analyze the relevance of these additional memory terms in two ways: First, we determine their magnitude and compare it to the magnitude of other terms in the GLE, and second, we interpret the GLE as a stochastic differential equation, perform a Markovian embedding and analyze the relevance of these terms for coarse-grained simulations. The results of these tests have direct implications for the validity of the Grote-Heynes theory.

The structure of the article is as follows: In \cref{sec:theoretical_foundation}, we introduce the specific projection operator that serves as the starting point for the analysis. From this, we show the exact form of the resulting GLE and some important properties of its terms. In \cref{sec:model_sys_and_sim}, the model system is introduced and details on the MD simulation procedure are given. The results of the numerical analysis of the MD simulations are given in \cref{sec:ana_I} and, the coarse-grained simulations setup and its subsequent analysis is discussed in \cref{sec:ana_II}. Finally, a conclusion is given in \cref{sec:conclusion}.

\section{Theoretical Foundation and Choice of the Projection Operator}\label{sec:theoretical_foundation}

We consider a canonical ensemble of a classical system. The entire set of coordinates that define a point in phase space are denoted as $\vec{\Gamma}$ and the phase-space density is $\rho(\vec{\Gamma})\sim\exp(H(\vec{\Gamma}))$ where $H(\vec{\Gamma})$ denotes the Hamiltonian of the system. We are interested in a coarse-grained model for one component of the relative displacement of two particles, e. g. the difference of the $x$-coordinate of the first and second particle $x:=r_{1,x}-r_{2,x}$. If both particles have the same mass $m$, we can define the $x$-component of the difference of their momenta as $p:=p_{1,x}-p_{2,x}=m\mathcal{L}x$ where $\mathcal{L}$ is the Liouvillian derived from $H(\vec{\Gamma})$.

Given a set of phase-space functions $\vec{A}(\vec{\Gamma})$, e.g. $\vec{A}=(p,x)^\top$, we can write the formal solution to the equations of motion as
\begin{align*}
    \dv{\vec{A}_t}{t} &= \mathcal{U}(t)\mathcal{L}\vec{A}
\end{align*}
where $\mathcal{U}(t)=\exp(t\mathcal{L})$ is the time-evolution operator. We use the subscript $t$ to denote the value at that time, and no subscript implies $t=0$, i.e., $\vec{A} = \vec{A}_0$. Next, we introduce an inner product via
\begin{align}
    \left( \vec{X}(\vec{\Gamma}), \vec{Y}(\vec{\Gamma}) \right)\! &:=\! \expval{\vec{X}(\vec{\Gamma}) \otimes \vec{Y}(\vec{\Gamma})}\! =\! \int\! \mathrm{d}\vec{\Gamma} \rho(\vec{\Gamma}) \vec{X}(\vec{\Gamma}) \otimes \vec{Y}(\vec{\Gamma}).
\end{align}
With this at hand, we can write the Mori projection operator \cite{Mori_Transport_1965} as
\begin{align}
    \mathcal{P}\ldots &= (\ldots,\vec{A})\cdot(\vec{A},\vec{A})^{-1}\cdot\vec{A}
\end{align}
and define its orthogonal projection $\mathcal{Q} \equiv 1 - \mathcal{P}$. Then the orthogonal dynamics are generated by the operator $\exp(t\mathcal{LQ})$ as shown in ref.~\cite{widder2025generalized}.

The corresponding generalized Langevin equation reads
\begin{align}
    \dv{\vec{A}_t}{t} &=\boldsymbol{\omega}\cdot\vec{A}_t - \int_0^t\dd\tau\,\mathbf{K}(t-\tau)\cdot\vec{A}_\tau+\vec{\eta}_t\label{eq:mori_gle}
\end{align}
where
\begin{align}
    \boldsymbol{\omega} &= (\mathcal{L}\vec{A},\vec{A})\cdot(\vec{A},\vec{A})^{-1}\\
    &=\begin{pmatrix}
        0 & \expval{x\mathcal{L}p}/\expval{x^2}\\
        1/m & 0
    \end{pmatrix}\,.
\end{align}

Similarly, one can show that only a single component of the memory-kernel matrix is nonzero (cf. \cref{app:mem_ker}), namely,
\begin{align}
    \boldsymbol{K}(t) &= \begin{pmatrix}
        K_{11}(t) & 0\\
        0 & 0
    \end{pmatrix}.
\end{align}
Together with the 2FDT,
\begin{align}
    \boldsymbol{K}(t)\cdot\expval{\vec{A}\otimes\vec{A}} &=\expval{\vec{\eta}_t\otimes\vec{\eta}_0},
\end{align}
\cref{eq:mori_gle} takes the very simple form of
\begin{align}
    \dv{p_t}{t} &= \frac{\expval{x\mathcal{L}p}}{\expval{x^2}} x_t -\int_0^t\dd\tau\,K_{11}(t-\tau)p_\tau+\eta_{1,t},\label{eq:gle_first_of_two}\\
    \dv{x_t}{t} &= \frac{1}{m}p_t.
\end{align}

To see how the first term on the right-hand side of \cref{eq:gle_first_of_two} may be related to a thermodynamic force, or more specifically a mean force, we introduce the probability distribution $P(x)$, i.e.~the probability to find the system having a position $x$ in the equilibrium ensemble. (In the literature on the projection-operator formalism this quantity is called "macroscopic probability density".)
The potential of mean force is then given by $\text{W}_\text{MF}(x)=-k_BT\ln(P(x))$ and the mean force is the derivative $F_{\text{MF}}=-\frac{\dd {\text{W}_\text{MF}}(x)}{\dd x}$. Note that this definition of the mean force is equivalent to defining it via the positional conditional expectation value of the force, $F_\text{MF}(x)=\text{E}[\mathcal{L}p|x]$. Using $P(x)$ and the conditional probability distribution $\rho(\vec{\Gamma})=P(x)\rho(\vec{\Gamma}|x)$, we obtain 
\begin{align}
    \expval{x\mathcal{L}p} &= \int\dd x\, P(x)xF_\text{MF}(x).
\end{align}
We see that a $\text{W}_\text{MF}$ of the shape $\text{W}_\text{MF}(x)=c_0+c_2 x^2$ that leads to a mean force $F_\text{MF}(x)=-2c_2 x$ will result in a mean-force term in the GLE because
\begin{align}
    \frac{\expval{x\mathcal{L}p}}{\expval{x^2}} x_t &= -2c_2\frac{\expval{x^2}}{\expval{x^2}}x_t\\
    &= F_\text{MF}(x_t).
\end{align}
For this special case of $\text{W}_\text{MF}(x)$ one actually obtains a GLE with a thermodynamic force, a single memory term linear in the momentum and a fluctuating force fulfilling a simple 2FDT.

Next, we raise the question: What is the simplest polynomial form of $\text{W}_\text{MF}(x)$ that does not yield such a simple description? The first idea might be to add a linear contribution to $\text{W}_\text{MF}(x)$. This, however, can easily be reduced to the case previously discussed by shifting $x$ appropriately. The next higher order of $x^3$ cannot be included as it would result in a thermodynamically unstable potential of mean force. Thus, we consider $\text{W}_\text{MF}(x)=c_0+c_2 x^2 + c_4 x^4$ the simplest nontrivial form. In this case the mean force is $F_\text{MF}(x)=-2c_2x-4c_4x^3$ and, thus, we choose $\vec{A}=(p, x, x^3)^\top$ as our set of observables in the Mori GLE (cf. \cref{eq:mori_gle}). For the sake of readability and since we do not vary the masses in this article, we set $m=1$ in the following. We find that
\begin{widetext}
    \begin{align}
        \boldsymbol{\omega} &= \begin{pmatrix}
        0 & -(p,p) & -(p,3x^2 p)\\
        (p,p) & 0 & 0\\
        (p,3x^2 p) & 0 & 0
        \end{pmatrix} \cdot
        \begin{pmatrix}
        (p,p) & 0 & 0\\
        0 & (x,x) & (x,x^3)\\
        0 & (x^3,x) & (x^3,x^3)
        \end{pmatrix} ^{-1}
    \end{align}
\end{widetext}
Therefore, the $\omega_{11}$ component is identically zero, and hence the equation of motion for the momentum is given by
\begin{align}
    \dv{p_t}{t} &=  \omega_{12}x_t + \omega_{13}x_t^3 -\int_0^t\dd \tau\, K_{11}(t-\tau)p_\tau\nonumber\\
    &\phantom{=} -\int_0^t\dd\tau\, \left[K_{12}(t-\tau)x_\tau +K_{13}(t-\tau)x_\tau^3\right]  + \eta_{1,t}\,.
    \label{eq:gle_momentum_component_triple_projection}
\end{align}
Note that the two positional memory kernels $K_{12}(t)$ and $K_{13}(t)$ are not independent of each other but satisfy the relation (derivation given in Appendix~\ref{app:mem_ker})
\begin{align}
    \frac{K_{12}(t)}{K_{13}(t)} &= \text{const}
    \label{eq:mk_ratio}
\end{align}
which can be used to test the numerically obtained kernels for consistency with the theory.
The equation of motion for the position is again 
\begin{align}
    \dv{x_t}{t} &= \frac{1}{m}p_t.
\end{align}
The third component of the evolution equation, the one for $x_t^3$, has a more complex structure. However, for all practical purposes, one may simply exploit the fact that $x_t^3$ follows straight-forwardly from $x_t$. \Cref{eq:gle_momentum_component_triple_projection} together with the two trivial equations for $x_t$ and $x_t^3$ are the basis for the analysis presented in the following.

This clearly differs from 
\begin{align}
    \dv{p_t}{t} &=  \omega_{12}x_t + \omega_{13}x_t^3 -\int_0^t\dd \tau\, K_{p}(t-\tau)p_\tau+ \eta_{t}\, ,
    \label{eq:gle_wrong_momentum_component}
\end{align}
i.e.~an equation with a mean force term, a singular memory kernel coupled to the momentum, and a fluctuating force satisfying a 2FDT with the kernel.

\begin{figure}
    \centering
    \includegraphics[width=\linewidth]{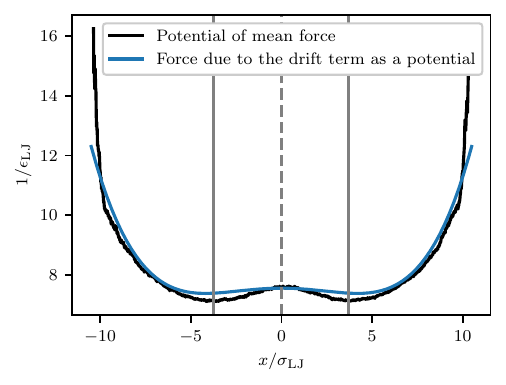}
    \caption{Potential of mean force calculated using the position distribution compared with the drift term integrated with time and shifted to match at $x = 0$. Vertical lines showcasing the positions considered for calculating passage times to cross the potential barrier (starting from either bold line and crossing the dashed line, i.e. changing sides)}
    \label{fig2}
\end{figure}

\begin{figure*}
    \centering
    \includegraphics[scale = 1]{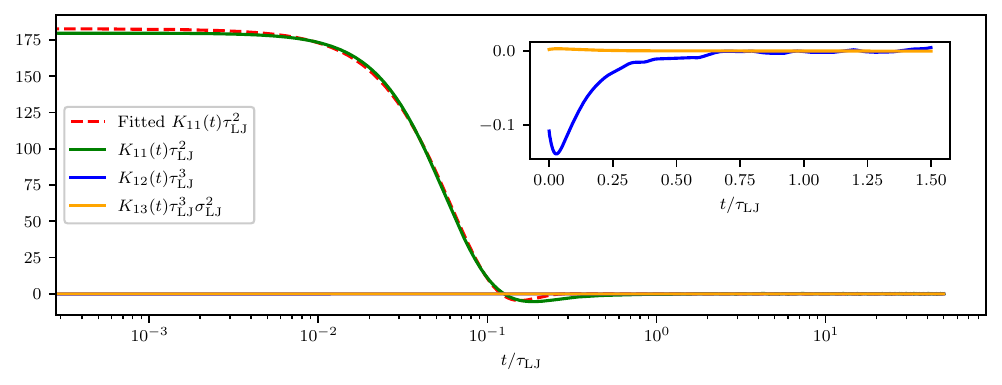}
    \caption{Memory kernel components as defined in \cref{eq:gle_momentum_component_triple_projection} along with the fitted function for $K_{11}(t)$ used for the coarse-grained simulations. Along with a zoomed in picture of the $K_{12}(t)$ and $K_{13}(t) \approx -K_{12}(t)/(40\sigma_\text{LJ}^2)$ components, showing that they are non-zero}
    \label{fig3}
\end{figure*}

\section{Model System and MD Simulations}\label{sec:model_sys_and_sim}
We want to investigate a system where the probability distribution and therefore the $\text{W}_\text{MF}$ follows the form highlighted in the above section, i.e., it is a function of $x^2$ and $x^4$. To achieve this in a nontrivial manner, we make the following ad-hoc model:

The system consists of 4096 particles, of which 4094 particles comprise the bath and interact through a Lennard-Jones (LJ) potential. The remaining two particles interact with the bath and with each other via the same LJ potential. In addition they have the same charge which results in a repulsion between them. The observed particles also have a harmonic potential from the center of the simulation box, and the coefficient of the spring force is homogeneous in the y and z-coordinates but differs in the x-coordinate. All of the particles have the same mass and are under the influence of an external quadratic potential near the walls that repels them from the edges of the simulation box. The simulation is carried out in a cubic box with a side length of $26\sigma_\text{LJ}$, where $\sigma_\text{LJ}$ and $\epsilon_\text{LJ}$ are the LJ diameter and energy respectively. The LJ cutoff for the interaction among the particles is set to $2\sigma_\text{LJ}$ and the LJ cutoff for the walls is $2.5\sigma_\text{LJ}$. The magnitude of the repulsive force is equal to $20\epsilon_\text{LJ}/\sigma_\text{LJ}$ when the particles are at a distance of $\sigma_\text{LJ}$ from each other. The force resulting from the harmonic potential from the center of the box on the observed particles is $0.1 \epsilon_\text{LJ}/\sigma_\text{LJ}$ along the x-axis at a distance of $\sigma_\text{LJ}$ along the x-axis from the center of the box and $1 \epsilon_\text{LJ}/\sigma_\text{LJ}$ along the y and z-axes at a distance of $\sigma_\text{LJ}$ along the respective axes from the center of the box. The harmonic potential near the walls has a coupling strength of $3.3\epsilon_\text{LJ}/\sigma_\text{LJ}^2$ starting from a distance of $3\sigma_\text{LJ}$ away from the edges of the box. Clearly, this model is not supposed to represent a realistic experimental system. 

The molecular dynamics simulations were performed using LAMMPS to propagate the system. The Verlet algorithm was used to integrate the system with a timestep of $\Delta t = 0.0005\tau_\text{LJ}$, where $\tau_\text{LJ}=\sqrt{\sigma_\text{LJ}^2/\epsilon_\text{LJ}}$ is the characteristic LJ time. The system is first brought to equilibrium using a Langevin thermostat, after which the system is propagated without a thermostat and the observables measured.

\section{Analysis I}\label{sec:ana_I}
Our first objective is to check whether we obtain a $\text{W}_\text{MF}$ of the desired form in the MD simulations. From \cref{fig2} we can see that this is indeed the case as we obtain a double-well potential.

The first check is to ensure that the additional memory kernels, which we do not expect to identically vanish, are, in fact, non-vanishing. This is confirmed by MD simulations and is shown in \cref{fig3}. We observe the $K_{11}(t)$ component to be an exponentially-decaying function. Surprisingly, the position coupled memory kernels do not have their maximum magnitude at $t=0$ as is the case with the momentum coupled component and what is intuitively expected. It should also be noted that $K_{12}(t)$ is primarily a negative-valued function, which seems unexpected at first glance, but the RHS of \cref{memory_components_ratio} is a negative quantity, implying that $K_{12}(t)$ or $K_{13}(t)$ must be negative.

We also numerically check whether we can reproduce the ratio of the components of the memory kernel (\cref{eq:mk_ratio}) highlighted in the previous section. The ratio is indeed a constant over time, having a value of about $K_{12}(t)/K_{13}(t)\approx-40 \sigma_\text{LJ}^2$.

Knowing each of the memory-kernel components and the momentum and position series from the simulations, we can check the average values of not just the kernels but instead each of the convolutions to give us an estimate of their relative magnitudes. We cannot directly compare the average values as they go to zero, so instead we compare the variances. After carrying out the analysis, we got the mean for each of these terms to be zero as expected, and their variances as

\begin{align}
    \text{E}\left[\left(\int_0^t\dd \tau\, K_{11}(t-\tau)p_\tau\right)^2\right] \approx 74 \sigma_\text{LJ}^2/\tau_\text{LJ}^4,
\end{align}
\begin{align}
    \text{E}\left[\left(\int_0^t\dd \tau\, K_{12}(t-\tau)x_\tau\right)^2\right] \approx 0.26 \sigma_\text{LJ}^2/\tau_\text{LJ}^4,
\end{align}
\begin{align}
    \text{E}\left[\left(\int_0^t\dd \tau\, K_{13}(t-\tau)x^3_\tau\right)^2\right] \approx 0.25 \sigma_\text{LJ}^2/\tau_\text{LJ}^4 ,
\end{align}
where the expectation values are calculated for all times greater than the decay time of the memory kernels and over all trajectories.

Looking at these values, the contribution of the momentum memory convolution is one order of magnitude greater than the position memory convolution, so it does not seem obvious that positional memory can be neglected in comparison. Also, the momentum is a fast fluctuating observable, and consequently so is the momentum coupled memory. In contrast, the position varies slowly, hence the positional memory behaves similarly to a longer-time push towards the center for small magnitudes of the x-coordinate (due to the negative sign of the $K_{12}(t)$ component, which can be seen in \cref{fig3}).

\section{Analysis II}\label{sec:ana_II}

In this section, we use stochastic interpretations of the two GLEs, the exact one (\cref{eq:gle_momentum_component_triple_projection}) and the one where the positional memory is neglected (\cref{eq:gle_wrong_momentum_component}), to generate new trajectories. This allows us to test what influence neglecting positional memory has on passage-time distributions. In order to run the coarse-grained simulations, we use the auxiliary-variables method due to its numerical simplicity and robustness. The central idea behind this approach is to come up with a simple Markovian embedding that automatically yields the correct combination of the momentum-memory term and the fluctuating force that fulfills the 2FDT \cite{Hauge_1973_Fluctuating,Shugard_1977_Dynamics,li2017computing,Doerries_2021_Correlation}.

In a first step, we fit the momentum-coupled memory kernel obtained from the MD simulations as an exponentially-decaying function to exactly follow the procedure outlined in ref.~\cite{li2017computing} (see \cref{fig3}). Following it, we fit the memory kernel to the form

\begin{align}
    k(t) &= \exp\left(-\frac{a}{2}t\right) \left[ \lambda_1 \cos{\omega t} + \lambda_2 \sin{\omega t} \right] ,
\end{align}

and obtain the parameters $a\approx46.4\tau_\text{LJ}^{-1}$, $\lambda_1 \approx 182.3\tau_\text{LJ}^{-2}$, $\lambda_2 \approx 197.9\tau_\text{LJ}^{-2}$ and $\omega \approx 19.8\tau_\text{LJ}^{-1}$.

The mean force and the positional memory terms of \cref{eq:gle_momentum_component_triple_projection} need to be added explicitly to the dynamics.

The shape of the $\text{W}_\text{MF}$ intuitively gives us a good physical quantity to measure whether the inexact GLE produces the same results as the exact GLE, that is, the passage time taken by the observed particles to cross the central barrier from either direction. That is, to start from either solid grey line in \cref{fig2} and cross the dashed grey line. In other words, start from either potential minimums and cross to the other side of the box. 

\begin{figure}
    \centering
    \includegraphics[width=\linewidth]{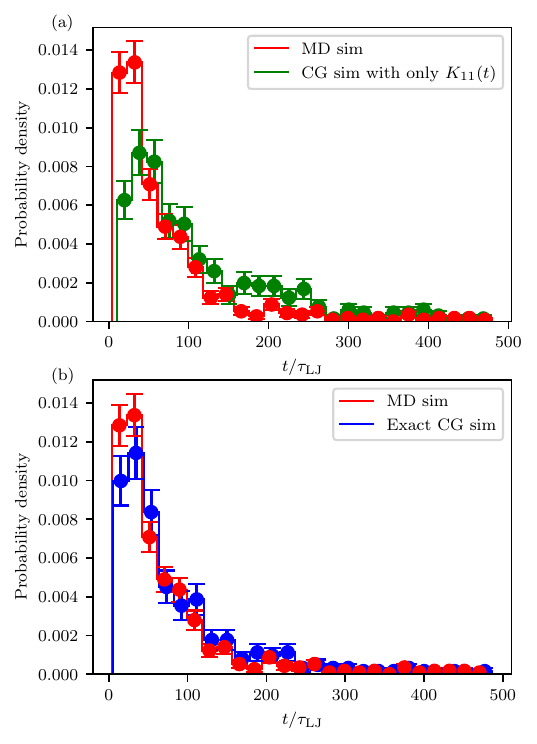}
    \caption{Passage time for MD simulations compared to coarse-grained simulations of (a) \cref{eq:gle_wrong_momentum_component} and (b) \cref{eq:gle_momentum_component_triple_projection}}
    \label{passage_times_comparison}
\end{figure}

The MD simulation data obtained are then analysed to generate the passage-time distribution. The resulting trajectories of the coarse-grained simulations are also analysed in the same way. As can be seen in \cref{passage_times_comparison}, neglecting positional memory yields a passage-time distribution that deviates strongly from the reference curve. In contrast, if one were to use the equation that includes positional memory instead, we would reproduce the results obtained from the MD simulations.

\section{Conclusion}\label{sec:conclusion}
In this article we analyzed the generalized Langevin equation which contains a quadratic potential of mean force (i.e.~a double well potential). Next to a memory term linear in the momentum there are memory terms in the position and in the third power of the position. We showed numerically that these additional memory terms contribute significantly to the evolution of the system. Neglecting these terms, as is often done when a generalised Langevin equation is used as an effective model for a complex process, affects the predictive power of the coarse-grained model. In particular the distribution of mean first passage times for barrier crossing depends on these terms.

%\TS{Some articles, in which double well potentials and GLEs were used to model something Catalysis in a biomolecular context: \cite{chaudhury2006approximate}, biomolecular folding \cite{satija2019generalized,singh2021generalized,roy2024memory}. An on the general properties of the GLE for barrier crossing in symmetric double-wells: \cite{schmidt2015simulation,drozdov2001improved,lavacchi2022non,zhou2025rapid}} Basic idea: Grote Hynes theory \cite{grote1980stable}, $\kappa_f$ from Grote Hynes is properly defined in ref.~\cite{Northrup_1980_stable} (by Northrup and Hynes)
\section*{Data Availability}
The data that supports the findings of this study are available from the corresponding author upon reasonable request.

\begin{acknowledgments}
The authors acknowledge funding by the Deutsche Forschungsgemeinschaft (DFG, German Research Foundation) Grant Nr. RTG 2717.
\end{acknowledgments}

\appendix

\section{Details on Memory Kernels}\label{app:mem_ker}
Formally, the memory-kernel matrix \cite{Mori_Transport_1965} is defined as
\begin{align}
    \boldsymbol{K}(t) &= \left(\exp(t\mathcal{LQ})\mathcal{L}\vec{A},\mathcal{QL}\vec{A}\right)\cdot\left(\vec{A},\vec{A}\right)^{-1}.
\end{align}
In the case of the two observables $\vec{A}=(p,x)^\top$, this matrix reads
\begin{align}
    \boldsymbol{K}(t) &= \begin{pmatrix}
        M_{11}(t) & 0\\
        0 & 0
    \end{pmatrix}\cdot\begin{pmatrix}
        1/(p,p) & 0\\
        0 & 1/(x,x)
    \end{pmatrix}
\end{align}
where $\mathcal{Q}x=0$ is used.

For $\vec{A}=(p,x,x^3)^\top$, one analogously finds
\begin{align}
    \boldsymbol{K}(t) &= \begin{pmatrix}
        M_{11}(t) & 0 & M_{13}(t)\\
        0 & 0 & 0\\
        M_{13}(-t) & 0 & M_{33}(t)
    \end{pmatrix}\cdot\begin{pmatrix}
        \beta/m & 0 & 0\\
        0 & a & b\\
        0 & b & c
    \end{pmatrix}
\end{align}
with
\begin{align}
    a&=\expval{x^6}/\left(\expval{x^2}\expval{x^6}-\expval{x^4}^2\right),\\
    b&=-\expval{x^4}/\left(\expval{x^2}\expval{x^6}-\expval{x^4}^2\right),\\
    c&=\expval{x^2}/\left(\expval{x^2}\expval{x^6}-\expval{x^4}^2\right).
\end{align}
This implies that
\begin{align}
    \frac{K_{12}(t)}{K_{13}(t)} &=\frac{b}{c}=-\frac{\expval{x^4}}{\expval{x^2}}=\text{const}.
    \label{memory_components_ratio}
\end{align}

\bibliography{literature}% Produces the bibliography via BibTeX.

\end{document}